\documentstyle[epsfig,12pt]{article}
\begin{document}
\title{An Observation of Autocorrelation of Wilson Loops on Lattice }
\author{{Da Qing Liu$^2$, Ying Chen$^2$ and Ji Min Wu$^{1,2}$}\\
        {\small $^1$CCAST(World Lab.), P. O. Pox 8730, Beijing, 100080,
          China}\\
        {\small $^2$Institute of High Energy Physics, Chinese
Academy of Sciences,}\\
        {\small P. O. Box 918-4, Beijing, 100039, P. R. China}}
\maketitle
\begin{center}
\begin{minipage}{5.5in}
\vskip 0.8in
{\bf Abstract}\\
An observation of autocorrelation of Wilson loops on lattice is presented,
especially for the small separation in Markov chain. We give a possible
explanation for such behavior. We also present the dependence
of autocorrelation behavior on the chosen operators in this paper.

\end{minipage}
\end{center}
\vskip 1in
\noindent

\newpage

\section {Introduction}
In order to extract a meaningful physical quantity, e.g. glueball mass,
from MC simulation of lattice gauge theory, we have to construct the
correlation function $G(t)$ for chosen operator
$o(\stackrel{\rightharpoonup}{x},t)$:
$$
G(t) =
\Sigma_{\stackrel{\rightharpoonup}{x}}
<o(\stackrel{\rightharpoonup}{x},t)o(\stackrel{\rightharpoonup}{0},0)>,
$$
to extract the glueball mass $m$ when $t \rightarrow \infty $,
$$
G(t) \propto e^{-mt}.
$$
It is the average over the independent configurations.
Meanwhile, we also have to
make correctly error estimate for the measurement. We should
generate a sequence of configurations and then choose samples separated by
enough sweeps to ensure their independence. Therefore it is needed to
investigate the correction from autocorrelation between configurations in
Markov chain. Denoting sequence in Markov chain by $T_{MC}$, some
authors$^{\cite{s1}}$ have studied autocorrelation with different algorithm
and regard  normalized autocorrelation asymptotic behavior with separation
$\Delta T_{MC}$ in Markov chain as $e^{-\Delta T_{MC}/\tau_{exp}}$. In order
to improve the fitting, ref. \cite{s2} redefined $\tau_{exp}$ via
$$\tau_{exp}=\limsup_{\Delta T_{MC}\rightarrow \infty}\frac{\Delta
T_{MC}}{-log \rho(\Delta T_{MC})},$$
where $$\rho(\Delta
T_{MC})=\frac{<v_{i}v_{i+\Delta T_{MC}}>-<v_i><v_{i+\Delta
T_{MC}}>}{<v_{i}v_{i}>-<v_i><v_i>}$$ is normalized autocorrelation and $v_i$
is the measured value of certain operator in the $i^{th}$ configuration. But
this formula always works at large $\Delta T_{MC}$ region$^{\cite{s2}}$.

Since autocorrelation contribution to error is mainly from small
$\Delta T_{MC}$ region, we investigates autocorrelation at small
$\Delta T_{MC}$ using MC simulation in this paper. From the simulation, we
find that autocorrelation behavior is like
$\rho \propto e^{-c\sqrt{\Delta T_{MC}}}$
rather than $e^{-c\Delta T_{MC}}$, especially at small $\Delta T_{MC}$ region.
This discrepancy is important since the main autocorrelation contribution to
error comes from small $\Delta T_{MC}$ region. In section 2, we give a possible
explanation for such behavior.

We also investigate the dependence of autocorrelation with the operators
chosen for extracting the glueball mass. It is found that autocorrelation
depends on the chosen operators and we also give a possible explanation of
this dependence in section 2. We present our simulation results in section
3 and the conclusion in section 4.

We adopt improved action on anisotropic lattice ( one can find the
notations and details in ref. \cite{s3}):

$$S_{II}=\beta\{ \frac{5 \Omega_{sp}}{3 \xi
u_s^4}+\frac{4 \xi \Omega_{tp}}{3u_s^2}-\frac{\Omega_{sr}}{12\xi
u_s^6}-\frac{\xi \Omega_{str}}{12u_s^4} \}.$$
To simplify, we restrict ourselves to make the simulation in SU(2) pure
gauge fields.

\section{An Explanation for the Autocorrelation Behavior at Small
$\Delta
T_{MC}$ Region}
A configuration can be described by $4N^4$ SU(2)
matrices, where $N^4$ is lattice site number. This means we can depict
configuration by $12N^4\stackrel{\bigtriangleup}{=}M$ variables, a factor of
3 multiplied here is due to 3-dimension Lie algebra for each link(We regard
it as $E^3$ space for simplification). Therefore, configuration can be
regarded as  one point $P$ in M-dimensional Euclidean space S.  To simplify,
we assume the spacing of  time-like and space-like links are of the same. Then
all the $M$ parameters, or coordinates, are of equivalence due to the
periodic boundary condition.

When updating configurations, point $P$ forms a trajectory
in S space. Define distance between two point
$x=(x_1,x_2,\cdots,x_M),y=(y_1,y_2,\cdots,y_M)\in S$ by
$$r(x,y)=\sqrt{(x_1-y_1)^2+(x_2-y_2)+\cdots+(x_M-y_M)^2}.$$
In one sample, we regard the same-time-slice  sub-configuration
as a point $P_1$ in $9N^3(\stackrel{\bigtriangleup}{=}M_1)$-dimensional
space $S_1$( There are N such sub-configurations in one sample. ).
As expected, the series of
$N$ points in one sample can  be regarded as the evolution of the
sub-configuration with Euclidean time $t$.  All the links in these sub-configurations
are space-like, and we can define distance $r'$ of
two points in $S_1$ in the similar way. After doing these, we find that point
$P_1$ in $S_1$ has two independent evolution behaviors: the first one is the
evolution with $T_{MC}$ at fixed $t$ and the second one is
the evolution with $t$ at fixed $T_{MC}$(in the same sample).

Suppose configurations  have approached to equilibrium after
enough preupdatings. Updating configuration $n$ times, we get representation
points $\{ x^{(i)} \}(i=1,2,\cdots,n)$ in $S$ space. We denote the
displacement of
two adjoining points by $\stackrel{\rightharpoonup}{r}^{(i)}=x^{(i+1)}-x^{(i)}.$
($ x^{(i)}$ is also a vector.) When $n\rightarrow \infty$, we get a
sequence of
$\stackrel{\rightharpoonup}{r}$ with certain distribution. Due to ergodicity
and periodic condition, it can be expected that the distribution of
$\stackrel{\rightharpoonup}{r}$ only depends on its modular value $r$. Since
the procedure of updating configuration, or Markov chain, forms a trajectory
in space $S$, and the procedure is of random, we may regard the
procedure as random  walking in space $S$. Then mean distance of two points
separated by $\Delta T_{MC}$ in Markov chain is given by $<r>=\sqrt{2D_0
\Delta T_{MC}},$ where $D_0$ is diffusion coefficient which only depends on the
choice of algorithm.

Now let's  consider sub-configuration evolves with $T_{MC}$
at fixed $t_1$ ( the sub-configuration consists of space-like links
in time-$t_1$-slice). We denote corresponding points and displacements in $S_1$
with $\{ {x'}^{(i)} \}$  and $\stackrel{\rightharpoonup}{r'}^{(i)}$. When $n$
is large enough, we expect that mean distance of two sub-configurations
adjoining in Markov chain is  $\sqrt{M_1/M}r_0$ where $r_0$ is mean distance
of two representation points in $S$ space adjoining in Markov chain. So, at
certain $t_1$, mean distance of two sub-configurations separated by $\Delta
T_{MC}$ in Markov chain is
\begin{equation}
<r'>=const \cdot \sqrt{\Delta T_{MC}}.
\end{equation}
The distribution of $\stackrel{\rightharpoonup}{r'}$ has no
dominant direction.

Then we consider sub-configuration evolves with time $t$ in one
configuration. When configuration approaches to the equilibrium, the most
possible contribution to configuration  is from the vicinity near
the solution of classical equation of motion, and we expect the distance of two
sub-configurations separated by time $\Delta t$ is a well-defined function. Up
to the first order, mean distance of two sub-configurations separated by
$\Delta t$ in the same sample in $S_1$ space can be expanded as
\begin{equation}
<r'>\simeq c'_1+c'_2\Delta t,
\end{equation}
where $c'_2>0$. The
distribution of $\stackrel{\rightharpoonup}{r'}$  has no
dominant direction, too. Eq.(2) can also be written as
\begin{equation}
\Delta t=c_1+c_2 <r'>.
\end{equation}

Since we are interested in operators for extracting glueball
mass, we will only consider the operator with certain quantum number $J^{PC}$.
Suppose $o$ is such kind of the operator and its measurement $o(t)$ in
sub-configuration is a function of point $P_1$ in $S_1$. Autocorrelation of
$o(t)$ is actually the correlation function $G(t)$. One can extract the
glueball  mass by its asympototic behavious
\begin{equation}
\rho(\Delta t)\propto e^{-m\Delta t}.
\end{equation}
On the other hand, due to Eq.(3), it is
\begin{equation}
\rho(\Delta t)\propto e^{-c_2 m <r'>}.
\end{equation}
When we calculate correlation
function with certain $\Delta T_{MC}$ or $\Delta t$, we should average it in
many samples, or average it in subspace $S_1$ due to ergodicity.
Therefore, $\rho$ is
function of $\stackrel{\rightharpoonup}{r'}$. Meanwhile, the distribution of
$\stackrel{\rightharpoonup}{r'}$ has no dominant direction with certain
$\Delta T_{MC}$ and $\Delta t$, so $\rho$ is only the function of $r'$.


Let us, now, consider the evolution of sub-configuration with $\Delta
T_{MC}$ again. Due to Eqs. (1), (5), we get

\begin{equation}
\rho(\Delta T_{MC})\propto e^{-mc_2\sqrt{\Delta T_{MC}}}.
\end{equation}
In our simulation, we find it is better to express the autocorrelation
$ \rho$ by
\begin{equation}
\rho(\Delta T_{MC})\propto{e^{-(c_0+c_2 m)\sqrt{\Delta T_{MC}}}}.
\end{equation}
Presumably, it is due to finite size effect in lattice simulation.
One can find
that autocorrelation gets weaker with
the increasing of the mass extracted from the correlation function of the
operator  $G(t) (c_2>0)$ .

\section {Simulation Results}
As we know, one can construct certain combination of Wilson loops to extract
the  mass of $J^{PC}$ glueball states. These special combination of Wilson
loops transforms according to certain representation of cubic
group$(A_1,~A_2,~E,~T_1,~T_2)^{\cite{s7} \cite{s8}}$. We choose some of these
operators to observe the autocorrelations. These operatore are the
combination of the certain prototype Wilson loop which are shown
in Fig. 1.
Operator a) belongs to representation $A_1^+(J^P=0^+)$; b) belongs to
representation $E^+(J^P=2^+)$; c) belongs to representation $T_2^+(J^P=2^+)$
and d) belongs to representation $A_1^-(J^P=0^-)$.
The prototype of operators a) and b) are plequettes. We choose different
combination coefficent and different orientation to make operator a)
belongs to representation $A_1^+$ and operator b) belongs to
representation
 $E^+(J^P=2^+)$.
Therefore, one can extract different mass states from them.
Operator b) and c) belong to different representations
of cubic group but correspond to the same continuum $J^P$ states ( Masses
extracted from
operator b) and c) approach to the same value in the continuum limit).

Our simulation is performed on a $8^3\times 24$ anisotropic lattice with
$\xi=\frac{a_s}{a_t}=3.0$ and $\beta=1.0$. The mass $m a_t$ extract from
$A_1^+,~E^+,~T_2^+,~A_1^-$ operators are 0.74(2),1.27(10),1.21(7), and 1.59(7)
respectively.

We present the operators autocorrelation behavior and their fitting curve
in Fig. 2. Except at the large $\Delta T_{MC}$ region where the noise will overwhelm
the signal, the formula (7) works very well.

When analyzing autocorrelation data for operator $P_{SS}$, spatial
plaquette
and for operator $R_{SS}$, spatial $2\times 1$ rectangle from ref.
\cite{s5}, one
also find our formula (7) does work well. Our argument is also
supported by the data for different operators of ref. \cite{s2} ( their
Fig.2),
although their data is to describe fermions.

When $\Delta T_{MC}$ is not small, the space $S$ can not be
regarded as Euclidean one, so Eq.(1) is not valid. At the same
time,
the expansion in Eq.(2) is doubtful at large $\Delta t$. Therefore,
the explanation in section 2 is not advisable at large $\Delta T_{MC}$ region or weak
autocorrelation region. One may expect that when $\Delta T_{MC}$ is very large,
the autocorrelation behavior will transform  into the behavior
$e^{-\Delta T_{MC}/ \tau_{exp}}$. This can be seen obviously in Fig. 2 a).
( To guide our eyes, we extract it in Fig.3 with $-ln \rho $ verse
$\Delta T_{MC}$ ).  We find that the fitting curve is
$-ln \rho= -0.0786+0.675\sqrt{\Delta T_{MC}}$ ( for small $\Delta
T_{MC}$) and
$-ln \rho= -2.01+0.261\Delta T_{MC}$ (for large $\Delta T_{MC}$)
respectively.
But this transition does not happen suddenly. In fact,
at intermediate $\Delta T_{MC}$ region, the formula
$-ln \rho=-1.0+0.104\Delta T_{MC}$ works as well as the first formula.

The dependence of autocorrelation on the mass $m a_t$ extracted from the
correlation function $G(t)$
is very intricacy. Fitting autocorrelation by
$\rho\propto e^{-c\sqrt{\Delta T_{MC}}}$, Fig.4 presents c
verse $m a_t$ by $ c =-0.65+1.94 m a_t$. Our
data implies that with the increasing of the mass, the autocorrelation gets
more weak, as expected in eq.(7).

\section {Conclusion}

From our simulation, one can find that autocorrelation behavior is like
$e^{-c\sqrt{\Delta T_{MC}}}$ at small
$\Delta T_{MC}$ region. As $\Delta T_{MC}$ increases, autocorrelation
approaches to $e^{-c\Delta T_{MC}}$ gradually, as one expected.
Therefore, we can get fair error estimation from the formula
$e^{-c\sqrt{\Delta T_{MC}}}$. An explanation is also given in this paper
for such behavior.

Our simulation also show, the autocorrelation gets more weak with the
increasing of the extracted mass $m a_t$. It is consistent with the fact,
as one usually expect, the
autocorrelation is dependent on the operator chosen.
Since $0^{++}$ glueball is the lightest state of glueballs, therefore, the
correction from autocorrelation is larger in $0^{++}$ glueball case then
the others.

\newpage
\section * {Figure Captions}
\begin{itemize}
\item
{\bf Figure 1}~~Prototypes of the corresponding operators a) - d) in the
paper.
Each operator is the different combination of the
prototype$^{\cite{s7}\cite{s8}}$ with directions in 3-dimensional space.
a) and b) are both plaquette, but we
choose different combination coefficients to make operator a)
 belong to representation $A_1^+$ and operator b) belong to $E^+$ in the
cubic group. We also choose proper coefficients to make operator c) belong
to $T_2^+$ and operator d) belong to $A_1^-$ in the cubic group.

\item {\bf Figure 2}~~autocorrelation behaviors $\rho$ of operator a) - d) in Fig. 1 and
their fitting curves. Fitting
curves(dashing curve):  a) is $\rho=e^{0.0786-0.675\sqrt{\Delta T_{MC}}}$; b) is
$\rho=e^{0.92-2.02\sqrt{\Delta T_{MC}}}$; c) is $\rho=e^{0.7-1.83\sqrt{\Delta
T_{MC}}}$; and d) is $\rho=e^{0.64-2.27\sqrt{\Delta T_{MC}}}$. The solid curves
in a) are $\rho=e^{2.01-0.261\Delta T_{MC}}$(large $\Delta T_{MC}$ region) and
$\rho=e^{-1.0-0.104\Delta T_{MC}}$(intermediate $\Delta T_{MC}$ region)
respectively. The vertical coordinate is $\rho$ and the horizontal one is
$\Delta T_{MC}$.

\item {\bf Figure 3}~~ $-ln\rho$ verse $m a_t$. The dashing curve is
$-ln\rho=-0.0786+0.675\sqrt{\Delta T_{MC}}$). The solid line are $-ln\rho=
-2.01+0.261\Delta T_{MC}$( large $\Delta T_{MC}$ region) and
$-ln\rho=1.0+0.104\Delta
T_{MC}$( intermediate $\Delta T_{MC}$ region) respectively.
The vertical coordinate is $-ln\rho$ and the horizontal one is the mass
extracted from the operator.

\item {\bf Figure 4} ~~ $c$ verse $m a_t$. The fitting line is
$c=-0.65+1.94m a_t$( The dashing line is $c=1.487m a_t$).
The vertical coordinate is $c$ and the horizontal one is the mass extracted from
the operator.
\end{itemize}


\begin{thebibliography}{99}
\bibitem{s1}
For instance, Montvay, Quantum Fields on a Lattice, Cambridge University
Press. 1994,
\bibitem{s2}
SESAM+T$\chi$L-Collaboration, Nucl. Phys. {\bf B}(Proc. Suppl.)63(1998) 946,
\bibitem{s3}
C. J. Morningstar and M. Peardon, Phys. Rev. {\bf D}56(1997) 4043,
\bibitem{s4}
J. B. Zhang, M. Jin and D. R. Ji, Chin. Phys. Lett. 15(1998) 865,
\bibitem{s5}
T. Draper, C. Nenkov and M. Peardon,
Nucl. Phys, {\bf B}(Proc. Suppl.)53(1997) 997,
\bibitem{s7}
B. Berg and A. Billoire, Nucl. Phys. {\bf B}221(1983) 109,
\bibitem{s8}
D. Q. Liu, J. M. Wu and Y. Chen, hep-lat/0011087.
\end{thebibliography}
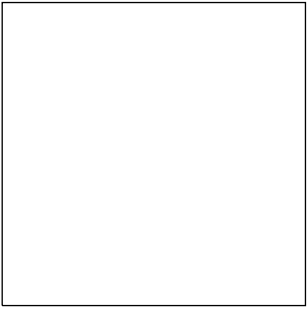
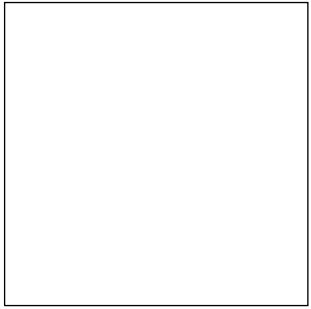
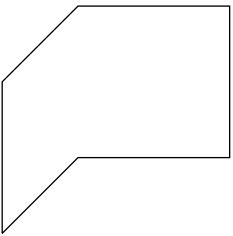
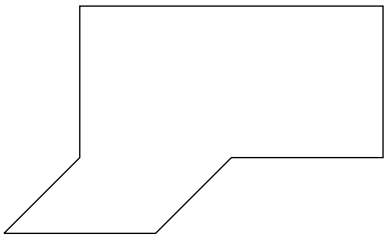
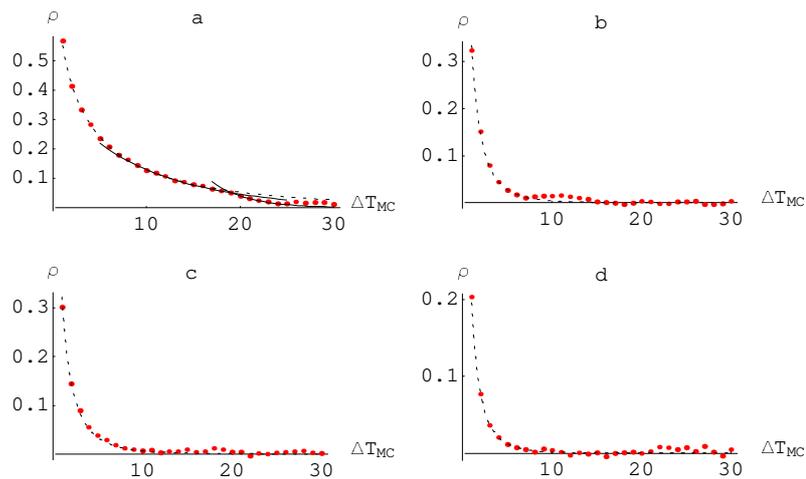
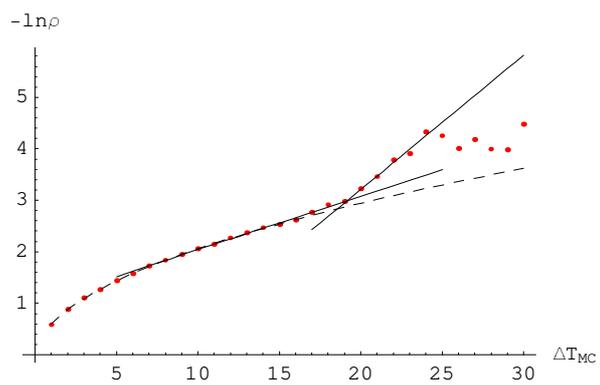
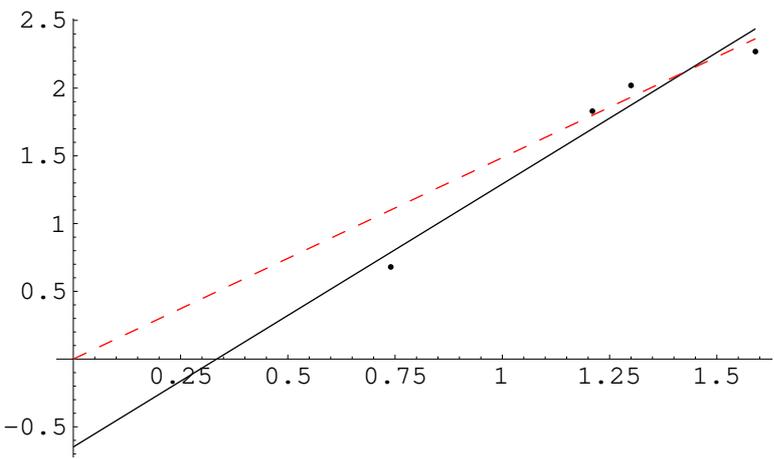
\end{document}